\documentstyle [aps,prl,preprint,floats,epsf]{revtex}

\draft

\begin{document}

\preprint{\tighten\vbox{\hbox{\hfil CLNS 97/1496}
                        \hbox{\hfil CLEO CONF 97-11}
                        \hbox{\hfil EPS Abstract 356}
}}

\title{Observation of the Dynamic Beta Effect at CESR with CLEO}

\date{\today}

\maketitle

\tighten

\begin{abstract}
	Using the silicon strip detector of the CLEO experiment
operating at the Cornell Electron-positron Storage Ring (CESR), we
have observed that the horizontal size of the luminous region
decreases in the presence of the beam-beam interaction from what is
expected without the beam-beam interaction.
The dependence on the bunch current agrees with the prediction
of the dynamic beta effect.  This is the first direct observation
of the effect.
\end{abstract}


\begin{center}

D.~Cinabro\\
{\it Wayne State University, Detroit, Michigan 48202}\\
\bigskip
M.~Chadha, S.~Chan, C.~O'Grady, J.S.~Miller, J.~Urheim,
A.~J.~Weinstein, and F.~W\"urthwein\\
{\it California Institute of Technology, Pasadena, California 91125}\\
\bigskip
S.~Prell, M.~Sivertz, and V.~Sharma\\
{\it University of California, San Diego, La Jolla, California 92093}
% ...

\end{center}

\newpage


	In a high energy colliding beam storage ring when two beams of
oppositely charged particles pass through each other there is an attractive
force between them.  Under certain circumstances this beam-beam interaction
can lead to a reduction in the transverse beam size at the interaction
point.  This ``dynamic beta'' effect is potentially important since a smaller
beam size corresponds to a higher luminosity.

	The dynamic beta effect has been long predicted~\cite{dybeta}.
It has been observed indirectly through its effect on luminosity at
CESR~\cite{CESR_dybeta}.  This letter
describes the first direct observation of the reduction in beam
size due to the effect.  The dynamic beta effect is
analyzed by modeling the beam-beam interaction as a linear focusing lens.
This focusing alters the beta, $\beta$, around the ring.  If the tune
is just above the half integer resonance there is a sharp
reduction in $\beta^\ast$, the beta at the interaction point.
This decreases the size of a single beam at the interaction point,
given by $\sigma^\ast = \sqrt{\beta^\ast \epsilon}$ where $\epsilon$ is
the emittance.  The strength of the linear lens and therefore the
reduction in $\beta^\ast$ becomes larger with increases in the bunch
current.

	CESR~\cite{CESR}
is a particularly good place to observe the dynamic beta effect.
The bunch current
is large, over 8mA, generating a large beam-beam interaction.
There is also a large range of bunch currents during normal CESR
operations as the beam lifetime is about three hours and the typical
fill lasts 75 minutes.  The collisions studied here have bunch currents
ranging from 4 to 8mA\@.
In the horizontal direction the
CESR lattice has a tune just above the half integer
contributing to a large dynamic beta effect.  In the vertical direction
the tune is not close to the half integer and thus bunch current
dependent effects are negligible on the vertical size of the beam.


	The CLEO detector has been described in detail elsewhere~\cite{CLEO}. 
A recent addition is a silicon strip vertex detector~\cite{SVX} which
is crucial
to the observation of the dynamic beta effect.  This consists of three layers
of silicon wafers arrayed in an octagonal geometry around the interaction
point.  The first measurement layer is at a radius of 2.3cm and the wafers
are read out on both sides by strips which are perpendicular to each
other.  The readout
strips have a pitch of about 100$\mu$m and with charge sharing the detector
has an intrinsic per point resolution of better than 20$\mu$m in both the
plane transverse to the beam, the $xy$ plane, and in the direction parallel
to the colliding beams, called $z$.


	We observe the luminous region with events of the type
$e^+e^- \to $ hadrons, which are defined as events
with more than two charged tracks and four or more significant energy deposits
in the electromagnetic calorimeter.  All the data are taken with 
electron-positron collisions with a center of mass energy equal or just
below the mass of the $\Upsilon$(4S) resonance.
We try to form a vertex from all
the tracks in the event that have hits in at least two of the three
layers of the silicon in both the $xy$-plane and $z$ direction. 
Those tracks with
a contribution to the vertex chi-square of over two per degree of freedom
are removed as being unlikely to share a common vertex with the other tracks.
This procedure is iterated until all such tracks are removed.
Finally the probability of the vertex chi-squared must be greater than
10\%.  Vertices formed from two or more tracks give three dimensional
points in space that are used to
determine the size of the luminous region.  A total of 77327 such
vertices are found.

	For each discrete fill of CESR the average position is found in the
vertical and horizontal directions.  
There are occasional large shifts in this average fill position that result
from tuning CESR for backgrounds or luminosity with the introduction
of closed orbit distortions through the interaction region.
These fill averages are subtracted from
all the points of the fill to correct for these shifts in the average
position of luminous region.  We have looked for changes in the average
beam position during the course of a fill and can discern none.  

	The distribution of these
positions in the horizontal ($x$) and vertical ($y$) directions is binned
and then fit
to a flat function, to account for vertices not resulting from beam-beam
collisions, plus a Gaussian shape.
The width of this Gaussian is taken as the observed width of the
luminous region.

	Another useful feature of the CESR lattice for the observation of the
dynamic beta effect is the disparity of the beam size between the
horizontal and vertical directions.  Without considering the effect
of the beam-beam interaction, the horizontal beta at the
interaction point, $\beta_{0x}^\ast$,
is 1.38m and the horizontal emittance,
$\epsilon_x$, is $2.0 \times 10^{-7}$m, giving
a Gaussian width of the beam in the
horizontal direction of $\sigma_{0x}^\ast = 525\mu$m.
In the vertical direction 
$\sigma_{0y}^\ast \approx 10\mu$m.
Note that when the two beams collide they create a luminous region which is
the overlap of the two colliding beams.  The size of this luminous
region is a factor of $\sqrt{2}$ smaller than the single beam size.

\begin{figure}
\begin{picture}(450,450)(0,0)
\put(0,25){\epsfxsize=6.0in\epsffile{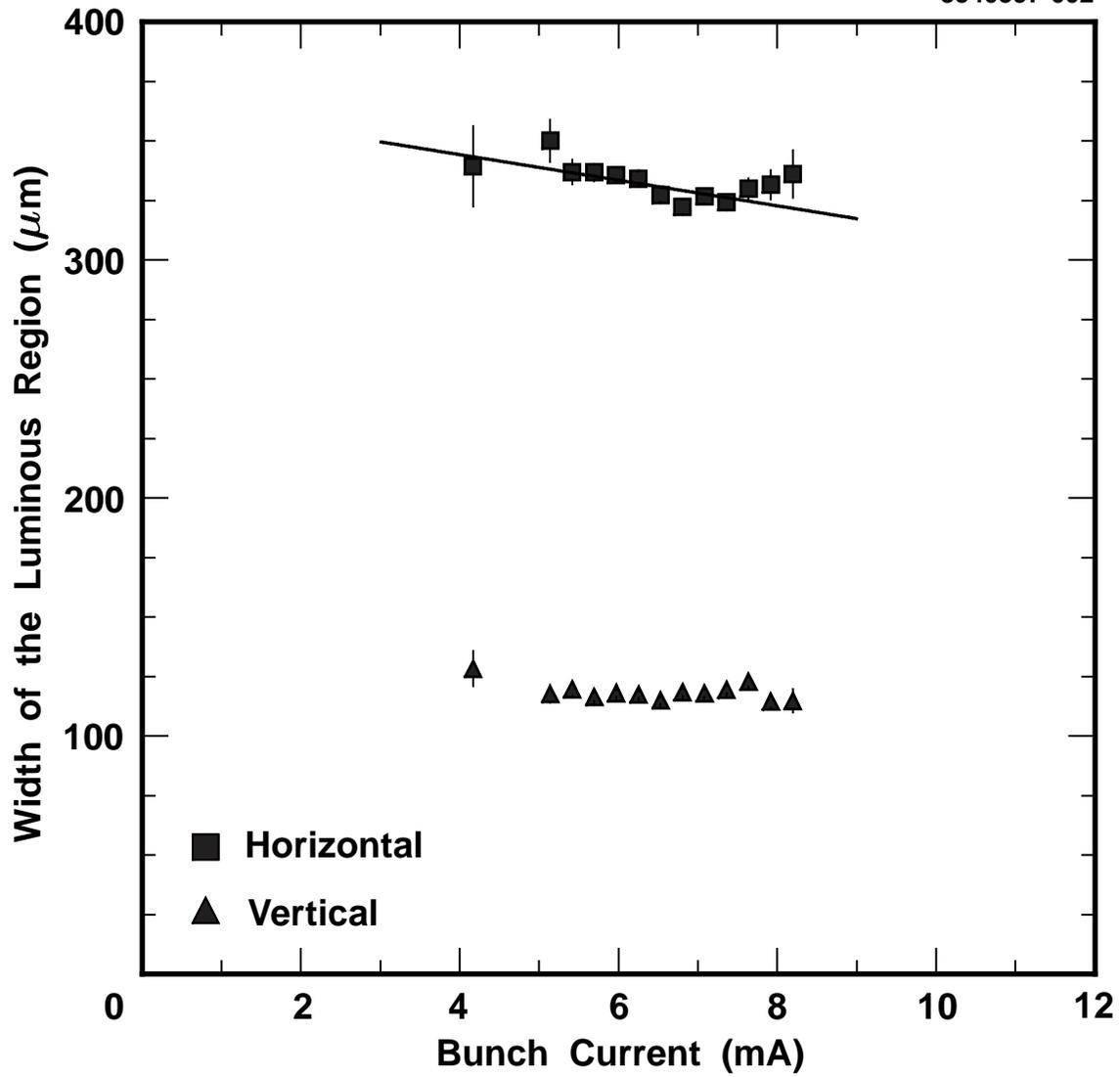}}
\end{picture}
\caption{The observed width of the luminous region in the horizontal and
vertical directions as function of the bunch current.  Also shown is the
line from the fit to the horizontal widths described in the text.} 
\label{fig:sigsvscur}
\end{figure}

	Figure~\ref{fig:sigsvscur} shows the observed width of the luminous
region in the horizontal and vertical directions as a function of the
bunch current.  The vertical width of around 120$\mu$m is much larger than the
expectation of $10/\sqrt{2} = 7\mu$m.  Studies with a Monte Carlo
based simulation
of hadronic events in the CLEO detector indicate that the vertical width
of the luminous region is a measure of the technique's resolution 
in the $xy$-plane.  
The observed horizontal width is
as small as 320$\mu$m which is smaller than the expected unperturbed
value of $525/\sqrt{2} = 370\mu$m.  If the observed vertical width
is interpreted
as the resolution these data imply an underlying horizontal width of
the luminous region of $\sigma_{x{\cal L}} = 295\mu$m which is
much below the expected unperturbed
value.  If we fit the observed widths as a function of the bunch
current to lines,
the vertical is seen to
be independent of the bunch current, while the horizontal width 
has a negative slope of $(-5.4 \pm 1.6)\mu$m/mA.  Both the observation
of a significantly smaller than expected horizontal width of the luminous
region in the presence of the beam-beam interaction and a significant
dependence of the horizontal width on the bunch
current are evidence of the dynamic beta effect.

\begin{figure}
\begin{picture}(450,450)(0,0)
\put(0,25){\epsfxsize=6.0in\epsffile{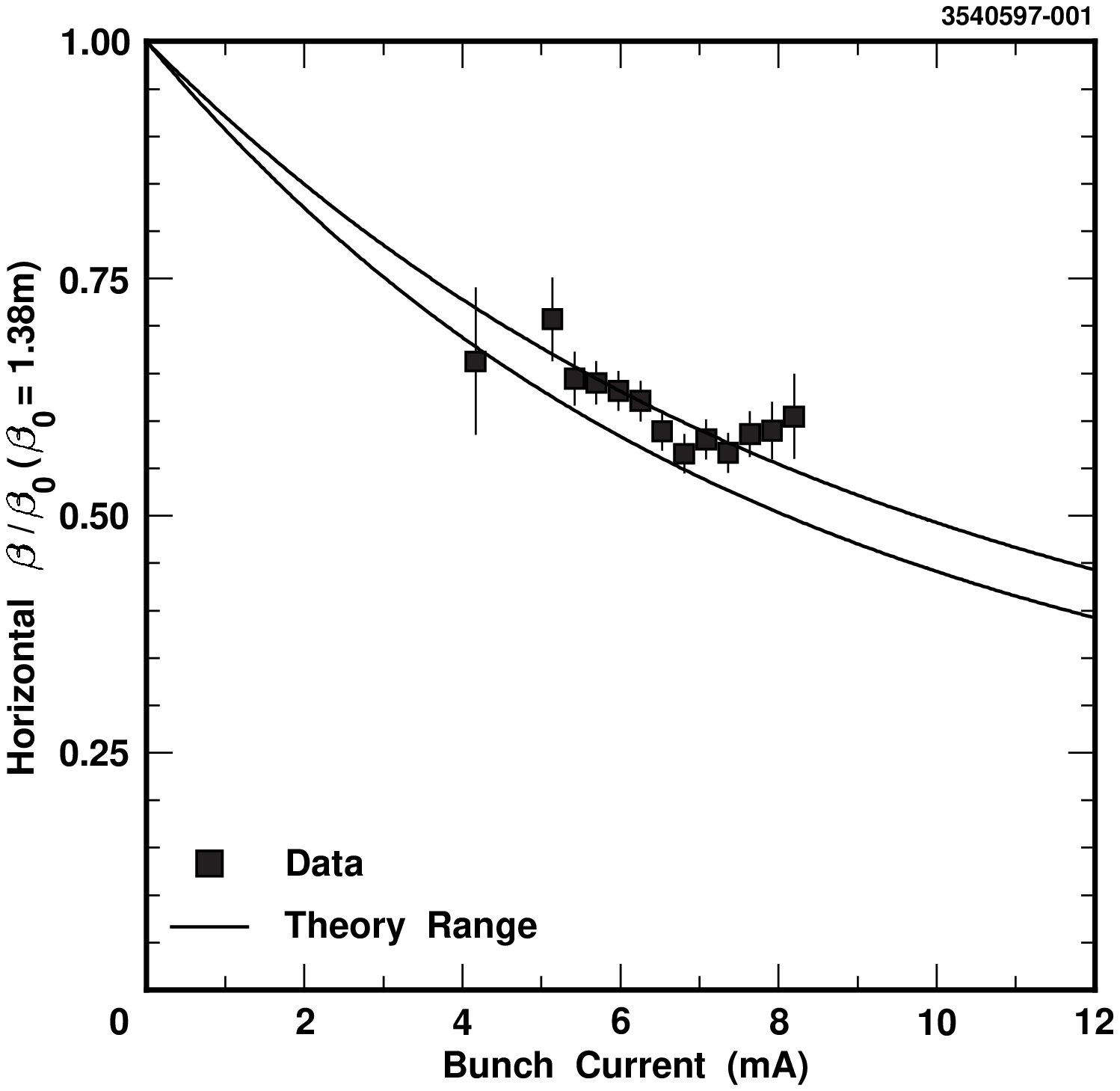}}
\end{picture}
\caption{The horizontal beta as function of bunch current.  Note that the
error on $\beta_{x0}^\ast$ of 5\% of its value is not included in the error
bars on the data.}
\label{fig:betavscur}
\end{figure}

	To compare with the prediction of the dynamic beta effect we
extract the horizontal beta from the width of the luminous region.
We use a resolution of $118 \pm 10\mu$m.  This is from a fit to
the vertical distribution independent of bunch current with the error
dominated by a 10$\mu$m jitter introduced by the correction for
the average fill position.  The horizontal emittance is taken from the
expectation of the CESR lattice.
It is parametrized by $\epsilon_x = 
(1.96 \times 10^{-7} + 1.46 \times 10^{-9} {\rm I_{bunch}} + 
5.15 \times 10^{-12} {\rm I_{bunch}^2})$m, where ${\rm I_{bunch}}$ is the
bunch current in milliAmps, which includes dynamic effects caused by
the beam-beam interaction.
We have measured the horizontal emittance using
$e^+e^- \to \mu^+\mu^-$ events observed with CLEO.  They give
$\epsilon_x = (1.87 \pm 0.65) \times 10^{-7}$m where the error is dominated
by the resolution on the angle between the two muons.
Then using $\beta_x^\ast = \sigma_{x{\cal L}}^2/2 \epsilon_x$
we calculate $\beta_x^\ast$ as a function of the bunch current as is shown in
Figure~\ref{fig:betavscur}.
The $\beta_{x0}^\ast$ of $(1.38 \pm 0.07)$m is taken from
a single beam measurement
using the observed phase advance through the interaction region.

	The theory curves shown in Figure~\ref{fig:betavscur} also include
the small effect of the beam-beam interaction at parasitic crossings away
from the interaction point where the beams are separated by a pretzel
orbit.  This effect serves to lower $\beta_x^\ast$ by about a 10\% 
at a bunch current of 12mA and is essentially linear from no effect at
zero bunch current.  The range of the theory comes from the
observed variation in the fractional integer tune from 0.537 to 0.544
from fill to fill.


	In conclusion, we have directly observed the dynamic beta effect
at CESR with
the silicon strip detector of the CLEO experiment.  We have seen that
the horizontal size of the luminous region is smaller
than expected when there
is a beam-beam interaction,
and that it decreases in size with increasing bunch current as the
dynamic beta effect predicts.
When we extract the horizontal beta as a function of the bunch
current it is seen to agree very well with
the expectation of the dynamic beta effect.  This is the first direct
observation of this effect.

\centerline{\bf ACKNOWLEDGEMENTS}
\smallskip
J.P.A. thanks                                           
the NYI program of the NSF, 
M.S. thanks the PFF program of the NSF,
M.S., H.N.N., and H.Y. thank the
OJI program of DOE, 
M.S. and V.S. thank the A.P. Sloan Foundation,
and M.S. thanks Research Corporation
for support.
This work was supported by the National Science Foundation, the
U.S. Department of Energy, and the Natural Sciences and Engineering Research 
Council of Canada.

%

\end{document}